\documentclass[12pt]{iopart}

\usepackage{iopams}  
\usepackage{color}
\usepackage{graphicx}
\usepackage{epsfig}
\usepackage{epstopdf}
\def\bm#1{\mbox{\boldmath $#1$}}
\begin{document}

\title[Ewald sums for inverse power-law interactions]{Ewald methods for inverse power-law interactions in tridimensional and quasi-two dimensional systems.}

\author{Martial MAZARS}
\address{Laboratoire de Physique Th\'eorique (UMR 8627),\\
Universit\'e de Paris Sud 11 et CNRS, B\^atiment 210, 91405 Orsay Cedex,
FRANCE\\[0.1in]
Preprint Number : LPT 10-66}
\ead{Martial.Mazars@th.u-psud.fr}
\begin{abstract}
In this paper, we derive the Ewald method for inverse power-law interactions in quasi-two dimensional systems. The derivation is done by using two different analytical methods. The first uses the Parry's limit, that considers the Ewald methods for quasi-two dimensional systems as a limit of the Ewald methods for tridimensional systems, the second uses Poisson-Jacobi identities for lattice sums. Taking into account the equivalence of both derivations, we obtain a new analytical Fourier transform intregral involving incomplete gamma function. Energies of the generalized restrictive primitive model of electrolytes ($\eta$-RPM) and of the generalized one component plasma model ($\eta$-OCP) are given for the tridimensional, quasi-two dimensional and monolayers systems. Few numerical results, using Monte-Carlo simulations, for $\eta$-RPM and $\eta$-OCP monolayers systems are reported.
\end{abstract}
\maketitle

\section{Introduction}

A quite general class of long range potentials is the inverse power law potentials. Generally, these potentials are defined as pseudo-potentials or as effective potentials ; the lattice sums with these potentials are used in solids state physics for the computation of structural integrals \cite{Misra:40,Born:43,Aschroft:book:76,Sugiyama:80,Sugiyama:84,Sugiyama:86,SmithAP:87,SmithAP:88} ; for instance, in the computation of the energy needed to permit the formation of an atomic vacancy in metals \cite{Sugiyama:86} or also to study the effect of a piezoelectric medium on structural properties of electronic bilayers in heterogeneous junctions AlGaAs-GaAs \cite{Fil:01}. The dependence of the liquid-gas transition for inverse power law interactions with $\eta=3+\sigma$ has also been studied in reference \cite{Camp:01}. Although, the analytical form of pseudo-potentials or effective potentials used are more complicated than a simple inverse power law, but rather like $f(r)/r^\eta$, in the following, we will restrict ourselves to $f(r)=1$.\\  
For a system with periodic boundary conditions in the three dimensions of the space, lattice sums of pseudo-potentials are 
\begin{equation}
\label{PowL_1}
\displaystyle \phi_\eta(\bm{r})=\sum_{\bm{L}_{\bm{n}}^{(d)}}\frac{1}{\mid\bm{r}+\bm{L}_{\bm{n}}^{(d)}\mid^\eta}
\end{equation}
with $\bm{L}_{\bm{n}}^{(d)}$ a symbolic notation for the $d$ dimensional periodic images of the basic cell that contains the system ; the basic cell is noted $\bm{L}_{\bm{0}}^{(d)}$ (or $\bm{L}_{\bm{0}}$ for tridimensional periodicities, or $\bm{S}_{\bm{0}}$ for two dimensional periodicities).\\
For $\eta=1$, this class of potential corresponds to the Coulomb interactions which lattice summations have been derived in various ways including Ewald methods \cite{DeLeeuw:80,Smith:08}, Lekner \cite{Lekner:91,Sperb:94,Mazars:01b} or plane-wise summations\cite{Sholl:67,Glasser:73}.\\
The case $\eta=3$ is also of a large interest in liquids and condensed matter physics, especially when dipolar interactions between particles are present in the systems \cite{DeLeeuw:80,Weis:book:05c}.\\
The limit $\eta\rightarrow\infty$ corresponds to the hard sphere model ; indeed, one has
\begin{equation}
\displaystyle \lim_{\eta\rightarrow \infty} \frac{1}{r^\eta} =+\infty \mbox{ for } r<1, \mbox{ and } \displaystyle \lim_{\eta\rightarrow \infty} \frac{1}{r^\eta}=0 \mbox{ for } r>1.
\end{equation}
The limit $\eta\rightarrow 0$ corresponds to an interaction potential that do not depend on the distance between the particles ($\lim_{\eta\rightarrow0} r^\eta =1$) ; thus, from the point of view of computer simulations, the limit $\eta\rightarrow 0$ corresponds to the ideal monoatomic gas. Logarithmic interactions or coulomb interaction in two dimensional systems cannot be obtained in the limit $\eta\rightarrow 0$ of inverse power law interactions of Eq.(\ref{PowL_1}). However, it may be obtained with the help of the relation
\begin{equation}
\label{PowL_log}
\lim_{\eta\rightarrow 0}\left[\frac{1}{\eta}\left(\frac{1}{r^\eta}-1\right)\right]=-\ln r
\end{equation}
The Ewald method for logarithmic interactions has been derived by Perram and de Leeuw \cite{Perram:81}. One has to take into account Eq.(\ref{PowL_log}) to recover their results from the Ewald sums for inverse power-law interactions developed in this paper.\\
Inverse power law interactions are also used in long-range dispersion force in Lennard-Jones fluids ($\eta=6$) ; some significant corrections are obtained by using Ewald summations in Lennard-Jones fluid rather than using a truncation of the potential \cite{Karasawa:89,Ou-Yang:05,Veld:07}.\\
The lattice sum (\ref{PowL_1}) for arbitrary $\eta$ has been obtained by Misra, Born and Bradburn \cite{Misra:40,Born:43} for tridimensional systems. The main purpose of the present paper is to compute this lattice sum for quasi-two dimensional systems (systems with spatial periodicities in only two directions of the space).\\
There are mainly two ways to obtain Ewald methods for quasi-two dimensional systems. First, one may obtain this formulation by taking the so-called Parry's limit \cite{Parry:75}; it consists in assuming that one of the spatial periodicity (say $L_z$) is extremely large compared to the other, the Ewald summations for tridimensional systems in the limit $L_z\rightarrow\infty$ are Ewald summations for quasi-two dimensional systems. This derivation has been applied successfully to Coulomb, Yukawa and dipolar interactions \cite{Weis:book:05c,Mazars:07d}. The second method consists in the application of Poisson-Jacobi identities (discrete Fourier transforms) to the two dimensional periodicities of the quasi-two dimensional systems. Both derivations are fully equivalent and they provide exactly the same results, this was shown extensively for Coulomb, Yukawa and dipolar interactions \cite{Mazars:07d,Mazars:07a,Weis:book:05c}.\\
Another method is based on a decomposition of the charge distributions with the help of different screening functions \cite{Fortuin:77,Rhee:89,Toukmaji:96,Lee:97,Salin:00,Johnson:07}. This method has been applied to coulomb interactions \cite{Fortuin:77,Rhee:89,Toukmaji:96,Lee:97}, to Yukawa potentials \cite{Salin:00} and inverse power-law interactions \cite{Johnson:07}. The screening of charges is frequently taken as a gaussian function, but this choice is not restrictive \cite{Fortuin:77,Rhee:89,Lee:97}.\\
The Poisson-Jacobi identity for lattice sums reads as
\begin{equation}
\label{Poisson_xD}
\displaystyle\sum_{\bm{L}_{\bm{n}}^{(d)}}f(\bm{r}+\bm{L}_{\bm{n}}^{(d)})=\frac{1}{V_d}\sum_{\bm{k}\in\mathcal{R}_d}\hat{f}\left(\frac{\bm{k}}{2\pi}\right)\exp\left(i\bm{k}.\bm{r}\right)
\end{equation}
where $V_d$ is the volume of the basic cell, $\mathcal{R}_d$ the reciprocal lattice associated with the lattice made of the periodic images of the system and $\hat{f}$ the Fourier transform of $f$ defined as 
\begin{equation}
\label{Fourier_d}
\displaystyle\hat{f}\left(\frac{\bm{k}}{2\pi}\right)=\int_{\mathbb{R}^d}f(\bm{x})\exp\left(-i\bm{k}.\bm{x}\right)d\bm{x}
\end{equation}
In this paper, we will derive Ewald summations for inverse power-law interactions in quasi-two dimensional systems by using both methods ; then, using the equivalence between both derivations, we will obtain a new analytical relation for a Fourier transform involving the incomplete gamma function. This Fourier transform is expressed in term of the generalized incomplete gamma function or  incomplete Bessel function \cite{Harris:08,Chaudhry:02}.\\
The paper is organised as follows. In the next section we give a short derivation of the Ewald method for the lattice summations with inverse power-law interactions in three dimensional systems. Then, using the result of section 2, in section 3 we derive the Ewald method for quasi-two dimensional systems by taking the Parry's limit and also by using the Poisson-Jacobi identity ; the equivalence between both derivations of the Ewald method for quasi-two dimensional systems then allows us to obtain a new Fourier transform for incomplete gamma function. In section 4, energies of the generalized restrictive primitive model of electrolytes ($\eta$-RPM) and of the generalized one component plasma model ($\eta$-OCP) are given for the tridimensional, quasi-two dimensional and monolayers systems. Several numerical results, using Monte-Carlo simulations for $\eta$-RPM and $\eta$-OCP monolayers systems are reported.

\section{Tridimensional systems.}

To apply the Ewald method to inverse power-law potentials, we begin as it is done for Coulomb interaction, by using the relation 
\begin{equation}
\label{PowL_2}
\displaystyle \frac{1}{\mid \bm{r}+\bm{L}_{\bm{n}}\mid^\eta} = \frac{1}{\Gamma\left(\frac{\eta}{2}\right)}\int_0^{\infty}\frac{dt}{t^{(1-\frac{\eta}{2})}}\exp(-\mid \bm{r}+\bm{L}_{\bm{n}}\mid^2 t)
\end{equation}
then we split the lattice sum  into two summations using a convergence parameter $\alpha$. One of the summations is on the periodic images of the system and the other one is transformed by using Poisson-Jacobi identity, both being rapidly convergent. We have
\begin{equation}
\label{PowL_3}
\hspace{-1.in}\begin{array}{ll}
\displaystyle \phi_\eta(\bm{r}) &\displaystyle = \sum_{\mbox{\small $\bm{L}_{\bm{n}}$} }\frac{1}{\Gamma\left(\frac{\eta}{2}\right)}\int_0^{\infty}\frac{dt}{t^{(1-\frac{\eta}{2})}}e^{-\mid \bm{r}+\bm{L}_{\bm{n}}\mid^2 t}\\
&\\
&\displaystyle = \sum_{\mbox{\small $\bm{L}_{\bm{n}}$} }\frac{1}{\Gamma\left(\frac{\eta}{2}\right)}\int_{\alpha^2}^{\infty}\frac{dt}{t^{(1-\frac{\eta}{2})}}e^{-\mid \bm{r}+\bm{L}_{\bm{n}}\mid^2 t}+ \frac{1}{\Gamma\left(\frac{\eta}{2}\right)}\int_0^{\alpha^2}\frac{dt}{t^{(1-\frac{\eta}{2})}}\sum_{\mbox{\small $\bm{L}_{\bm{n}}$} }e^{-\mid \bm{r}+\bm{L}_{\bm{n}}\mid^2 t}
\end{array}
\end{equation}
For the second contribution, we use the Poisson-Jacobi relation written as
\begin{equation}
\label{PoissonJacobi3D}
\sum_{\mbox{\small $\bm{L}_{\bm{n}}$} }e^{-\mid \bm{r}+\mbox{\small $\bm{L}_{\bm{n}}$}\mid^2 t}=\frac{1}{V}\left(\frac{\pi}{t}\right)^{3/2}\sum_{\bm{k}\in\mathcal{R}}e^{j\bm{k}.\bm{r}}\exp\left(-\frac{k^2}{4}\frac{1}{t}\right)
\end{equation}
\begin{table}
\begin{center}
\footnotesize
\begin{tabular}{|c|ccc|}
\hline
\hline
&&&\\
Power &  Real Space                            & Reciprocal Space                   & Contributions \\
           &  Contributions $\Phi_R^{(3)}$ & Contributions $\Phi_k^{(3)}$  &  for $\bm{k}=0$ \\
&&&\\
\hline
&&&\\
$\eta=4$                 & $\displaystyle \frac{e^{-\alpha^2 r^2}}{r^4}\left(1+\alpha^2 r^2\right)$ & $\displaystyle \frac{\pi^2}{V} \left[-k\mbox{ erfc}\left(\frac{k}{2\alpha}\right)+\frac{2\alpha}{\sqrt{\pi}}e^{-k^2/4\alpha^2}\right]$ & $\displaystyle \frac{2\pi^{3/2}}{V}\alpha$ \\
&&&\\
$\eta=3$                 & $\displaystyle \frac{1}{r^3}\left[\mbox{ erfc}(\alpha r)+\frac{2\alpha r}{\sqrt{\pi}}e^{-\alpha^2 r^2}\right]$ & $\displaystyle -\frac{2\pi}{V}\mbox{ Ei}\left(-k^2/4\alpha^2\right)$ & $\displaystyle \frac{4\pi}{V} \ln \left(\frac{\alpha}{\epsilon}\right)$ \\
&&&\\
$\eta=2$                 & $\displaystyle \frac{e^{-\alpha^2 r^2}}{r^2}$ & $\displaystyle \frac{2\pi^2}{V}\frac{\mbox{ erfc}\left(k/2\alpha\right)}{k}$ & $\displaystyle \frac{2\pi^{3/2}}{V}\left(\frac{1}{\epsilon}-\frac{1}{\alpha}\right)$  \\
&&&\\
$\eta=1$                 & $\displaystyle\frac{\mbox{ erfc}(\alpha r)}{r} $ & $\displaystyle \frac{4\pi}{V}\frac{e^{-k^2/4\alpha^2}}{k^2} $ & $\displaystyle \frac{\pi}{V}\left(\frac{1}{\epsilon^2}-\frac{1}{\alpha^2}\right)$ \\
&&&\\
\hline
\hline
\end{tabular}
\end{center}
\caption[Contributions r\'eelles et r\'eciproques des sommes d'Ewald tridimensionnelles pour une interaction en loi de puissance.]{\small Analytical formulas for the real and reciprocal space contributions in tridimensional Ewald summations for inverse power law interactions $1/r^\eta$ for some integer values of $\eta$. $\Phi_R^{(3)}$ and $\Phi_k^{(3)}$ contributions are given by (\ref{Cont_eta}). For $\eta\leq 3$, the contribution for $\bm{k}=0$ has a diverging behaviour which asymptotic expansions are given by using the prescription $\epsilon$ defined in equation (\ref{PowL_41}).}
\label{Tab3D_Powlaw_1}
\normalsize
\end{table}
we found
\begin{equation}
\label{PowL_4}
\begin{array}{ll}
\displaystyle \phi_\eta(\bm{r})&\displaystyle =\sum_{\bm{L}_{\bm{n}}} \Phi_R^{(3)}\left(\eta,\alpha ; \mid\bm{r}+\bm{L}_{\bm{n}}\mid\right)+\sum_{\bm{k}\neq 0}\Phi_k^{(3)}(\eta,\alpha ; k)\exp\left( j\bm{k}.\bm{r} \right)\\
&\\
&\displaystyle +\frac{1}{\Gamma\left(\frac{\eta}{2}\right)}\frac{\pi^{3/2}}{V}\int_0^{\alpha^2}\frac{dt}{t^{(5-\eta)/2}}
\end{array}
\end{equation}
where the last term is the contribution for $\bm{k}=0$ (it diverges for $0<\eta\leq 3$ and it is finite for $\eta> 3$). For Coulomb interaction, the contribution for $\bm{k}=0$ is closely related to the {\it macroscopic boundary condition} : a supplemental boundary condition imposed very far from the basic cell \cite{Smith:08,DeLeeuw:80}. Without further specification, the $\bm{k}=0$ contribution is diverging ; this can be cancelled in coulomb interaction  by assuming electroneutrality of the system in the basic cell. Similarly, for inverse power-law potential this contribution is diverging if $\eta\leq d$ ($d=3$, in this section - in section 4, we show how electroneutrality of the system in the basic cell suppresses this divergence). The way by which this IR-divergence is cancelled for inverse power-law potentials depends on the particular physical situation and on systems one studies. This IR-divergence is outlined below by the introduction of the prescription $\epsilon$ as
\begin{equation}
\label{PowL_41}
\displaystyle \int_0^{\alpha^2}\frac{dt}{t^{(5-\eta)/2}}=\lim_{\epsilon\rightarrow 0}\int_{\epsilon^2}^{\alpha^2}\frac{dt}{t^{(5-\eta)/2}}
\end{equation}
that gives the asymptotic expansion of the last integral in Eq.(\ref{PowL_4}).\\ 
Both functions $\Phi_R$ and $\Phi_k$ are given by 
\begin{equation}
\label{Cont_eta}
\left\{\begin{array}{ll}
\displaystyle \Phi_R^{(3)}(\eta,\alpha ; r)&\displaystyle =\frac{\Gamma\left(\frac{\eta}{2},\alpha^2 r^2\right)}{\Gamma\left(\frac{\eta}{2}\right)r^\eta}\\
&\\
\displaystyle \Phi_k^{(3)}(\eta,\alpha ; k)&\displaystyle=\frac{\pi^{3/2}}{V}\left(\frac{4}{k^2}\right)^{\frac{(3-\eta)}{2}}\frac{\Gamma\left(\frac{3-\eta}{2},\frac{k^2}{4\alpha^2}\right)}{\Gamma\left(\frac{\eta}{2}\right)}
\end{array}
\right.
\end{equation}
with $\Gamma(a,z)$ the complementary incomplete gamma function. These results agree with the derivations done in refs.\cite{Karasawa:89,Ou-Yang:05} ; Table \ref{Tab3D_Powlaw_1} gives some analytical forms of real and reciprocal contributions for few integer values of $\eta$.

\section{Quasi-two dimensional systems.}

To compute surface properties based on effective or pseudo-potentials, Ewald summations for quasi-two dimensional systems are of interest \cite{Fil:01}. Quasi-two dimensional systems are heterogeneous systems with some anisotropies in their spatial extensions, their numerical studies are done with partial boundary conditions : periodic boundary conditions are taken in directions with large spatial extensions and other boundary conditions are taken in directions with smaller extensions. For these systems, the lattice sums of inverse power law interactions are given by
\begin{equation}
\label{PowL_5}
\displaystyle \phi_\eta(\bm{r}) =\sum_{\bm{S}_{\bm{n}}}\frac{1}{\mid\bm{r}+\bm{S}_{\bm{n}}\mid^\eta}
\end{equation}
with $\bm{S}_{\bm{n}}$ ($\equiv \bm{L}_{\bm{n}}^{(2)}$) the symbolic notation for periodic images of the system in the 2D geometry.\\
As stated in the introduction, there are mainly two analytical ways to derive the Ewald method for a quasi-two dimensional system with a given interaction. First, if one already knows the analytical form of the Ewald summations for the corresponding full periodic tridimensional system, then one may make particular  one direction of the space ($Oz$ axis for instance) along which no periodic image is taken ; this is achieved by taking the limit $L_z\rightarrow \infty$ with $L_z$ the periodicity in the $z$ direction of the tridimensional system. This derivation was obtained by Parry for Coulomb interactions \cite{Parry:75} and also, more recently, for Yukawa interaction \cite{Mazars:07d} ; it can also be done for dipolar interactions. For systems in the Parry's limit, we use the notations
\begin{equation}
\label{Para_2D}
V=L_zA \mbox{            }  ; \mbox{            }  \bm{r}=\bm{s}+z\hat{\bm{e}}_z \mbox{            }  ;  \mbox{            }  \bm{L}_{\bm{n}}=\bm{S}_{\bm{n}}+n_zL_z \hat{\bm{e}}_z \mbox{ and }  \bm{k}=\bm{G}+k_z\hat{\bm{e}}_z        
\end{equation}
with the component $k_z$ of the wave vectors in the $Oz$-direction written as $k_z=2\pi n_z/L_z$ with $n_z$ integer and the two dimensional wave vectors $\bm{G}$ belonging to the reciprocal lattice associated with the two dimensional lattice defined by the periodic boundary conditions.\\
When the limit $L_z\rightarrow\infty$ is taken, the reciprocal space contribution has to be considered with caution. According to notations of equation (\ref{Para_2D}), one has to separate the summations on the wave vectors as 
\begin{equation}
\label{Decompo_2D}
\begin{array}{ll}
\displaystyle \sum_{\bm{k}\neq0}\bullet=\sum_{\bm{G}}\sum_{k_z}\mbox{}^{\prime}\bullet &\displaystyle : (i) \mbox{  if $\bm{G}\neq 0$, then $k_z = 0$ is allowed ;} \\ 
&\displaystyle (ii) \mbox{  if $\bm{G}= 0$, then $k_z \neq 0$}
\end{array}
\end{equation}
With these notations, the summations on the reciprocal vectors are split as
\begin{equation}
\label{2Dk_Parry}
\begin{array}{ll}
\hspace{-1.in}\displaystyle \sum_{\bm{k}\neq 0}\Phi_k^{(3)}(\eta,\alpha ; k)\exp\left( j\bm{k}.\bm{r} \right)&\displaystyle = \sum_{\bm{G}\neq 0} e^{j\bm{G}.\bm{s}}\left[\frac{1}{L_z}\sum_{n_z=-\infty}^{+\infty}\Phi_k^{(3)}(\eta,\alpha ; \sqrt{G^2+k_z^2})e^{jk_z z} \right]\\
&\\
\hspace{-1.in}&\displaystyle +T_{\bm{G}= 0}^{(\eta)}(\alpha,z)
\end{array}
\end{equation}
where the contribution $T_{\bm{G}= 0}^{(\eta)}$ corresponds to the case $(ii)$ of Eq.(\ref{Decompo_2D}), it is given by
\begin{equation}
\label{PowL_5b}
\displaystyle T_{\bm{G}= 0}^{(\eta)}(\alpha,z)= \frac{4^{(3-\eta)/2}\pi^{1/2}}{\Gamma(\frac{\eta}{2})}\frac{\pi}{A}\lim_{L_z\rightarrow\infty}\left[\frac{1}{L_z}\sum_{k_z\neq 0}\frac{\Gamma\left(\frac{3-\eta}{2},\frac{k_z^2}{4\alpha^2}\right)}{k_z^{(3-\eta)}}e^{j k_z z}\right]
\end{equation}
and it includes a non trivial contribution that depends on $z$ and also, if $\eta \leq 2$, it has an IR-divergence.\\
The summations over periodic image in the real space are also split as
\begin{equation}
\label{PowL_5c}
\begin{array}{ll}
\hspace{-1.in}\displaystyle \sum_{\bm{L}_{\bm{n}}} \Phi_R^{(3)}\left(\eta,\alpha ; \mid\bm{r}+\bm{L}_{\bm{n}}\mid\right)&\displaystyle = \sum_{\bm{S}_{\bm{n}}}\frac{\Gamma\left(\frac{\eta}{2},\alpha^2 \mid (\bm{s}+\bm{S}_{\bm{n}})+z\hat{\bm{e}}_z\mid^2 \right)}{\Gamma\left(\frac{\eta}{2}\right) \mid (\bm{s}+\bm{S}_{\bm{n}})+z\hat{\bm{e}}_z\mid^\eta}\\
&\\
\hspace{-1.in}&\displaystyle+\sum_{\bm{S}_{\bm{n}}}\sum_{n_z\neq 0}\frac{\Gamma\left(\frac{\eta}{2},\alpha^2 \mid (\bm{s}+\bm{S}_{\bm{n}})+(z+n_zL_z)\hat{\bm{e}}_z\mid^2 \right)}{\Gamma\left(\frac{\eta}{2}\right) \mid (\bm{s}+\bm{S}_{\bm{n}})+(z+n_zL_z)\hat{\bm{e}}_z\mid^\eta}
\end{array}
\end{equation}
In the Parry's limit ($L_z\rightarrow \infty$), only the first contribution in right hand side of Eq.(\ref{PowL_5c}) survives, then Eq.(\ref{PowL_5}) can be cast into the form
\begin{equation}
\label{PowL_6}
\hspace{-1.in}\displaystyle \phi_\eta(\bm{r})=\sum_{\bm{S}_{\bm{n}}} \Phi_R^{(3)}\left(\eta,\alpha ; \mid\bm{s}+\bm{S}_{\bm{n}}+z\hat{\bm{e}}_z\mid\right) +\sum_{\bm{G}\neq 0}\Phi_G^{(Q2)}(\eta,\alpha, z ; G)e^ {j\bm{G}.\bm{s}}+T_{\bm{G}= 0}^{(\eta)}(\alpha,z)
\end{equation}
where $\Phi_R^{(3)}(\eta,\alpha ; r)$ is given by Eq.(\ref{Cont_eta}) and $\Phi_G^{(Q2)}(\eta,\alpha, z ; G)$ is computed in the Parry's limit as
\begin{equation}
\label{PowL_7}
\begin{array}{ll}
\hspace{-1.in}\displaystyle \Phi_G^{(Q2)}(\eta,\alpha, z ; G) &\displaystyle = \frac{4^{(3-\eta)/2}\pi^{1/2}}{\Gamma(\frac{\eta}{2})}\frac{\pi}{A} \lim_{L_z\rightarrow\infty}\left[\frac{1}{L_z}\sum_{k_z\neq 0}\frac{\Gamma\left(\frac{3-\eta}{2},\frac{1}{4\alpha^2}(G^2+k_z^2)\right)}{(G^2+k_z^2)^{(3-\eta)/2}}e^{j k_z z}\right]\\
&\\
\hspace{-1.in}&\displaystyle = \frac{4^{(3-\eta)/2}\pi^{1/2}}{\Gamma(\frac{\eta}{2})}\frac{\pi}{A}\left[\frac{1}{2\pi}\int_{-\infty}^{+\infty}dk \frac{\Gamma\left(\frac{3-\eta}{2},\frac{1}{4\alpha^2}(G^2+k^2)\right)}{(G^2+k^2)^{(3-\eta)/2}}e^{j k z} \right]
\end{array}
\end{equation}
The second method to derive the Ewald method for quasi-two dimensional systems follows exactly the same derivation as the one done in section 2 for tridimensional systems. We begin by applying Eq.(\ref{PowL_2}) to the lattice sums (\ref{PowL_5}) and we split the integral into two contributions introducing $\alpha$. Then, we have
\begin{equation}
\label{PowL_7a}
\hspace{-1.in}\displaystyle \phi_\eta(\bm{r}) = \sum_{\mbox{\small $\bm{S}_{\bm{n}}$} }\frac{1}{\Gamma\left(\frac{\eta}{2}\right)}\int_{\alpha^2}^{\infty}\frac{dt}{t^{(1-\frac{\eta}{2})}}e^{-\mid \bm{r}+\bm{S}_{\bm{n}}\mid^2 t}+ \frac{1}{\Gamma\left(\frac{\eta}{2}\right)}\int_0^{\alpha^2}\frac{dt}{t^{(1-\frac{\eta}{2})}}e^{-z^2 t}\sum_{\mbox{\small $\bm{S}_{\bm{n}}$} }  e^{-\mid \bm{s}+\bm{S}_{\bm{n}}\mid^2 t}
\end{equation}
The Poisson-Jacobi identity in two dimensions, with $\mathcal{R}_2$ the reciprocal lattice, 
\begin{equation}
\label{PoissonJacobi2D}
\sum_{\mbox{\small $\bm{S}_{\bm{n}}$} }e^{-\mid \bm{s}+\mbox{\small $\bm{S}_{\bm{n}}$}\mid^2 t}=\frac{1}{A}\left(\frac{\pi}{t}\right)\sum_{\bm{G}\in\mathcal{R}_2}e^{j\bm{G}.\bm{s}}\exp\left(-\frac{G^2}{4}\frac{1}{t}\right)
\end{equation}
is applied to the second contribution, thus we obtain
\begin{equation}
\label{PowL_7b}
\hspace{-1.in}\begin{array}{ll}
\displaystyle \phi_\eta(\bm{r}) &\displaystyle = \sum_{\mbox{\small $\bm{S}_{\bm{n}}$} }\Phi_R^{(3)}\left(\eta,\alpha ; \mid\bm{r}+\bm{S}_{\bm{n}}\mid\right)+\frac{1}{\Gamma(\frac{\eta}{2})}\frac{\pi}{A}\sum_{\bm{G}\neq 0}e^{j \bm{G}.\bm{s}}\int_0^{\alpha^2}\frac{dt}{t^{(2-\frac{\eta}{2})}}e^{-z^2 t-\frac{G^2}{4t}}\\
&\\
&\displaystyle+\frac{1}{\Gamma(\frac{\eta}{2})}\frac{\pi}{A}\int_0^{\alpha^2} \frac{dt}{t^{(2-\frac{\eta}{2})}}e^{-z^2 t}
\end{array}
\end{equation}
The last contribution in Eq.(\ref{PowL_7b}) is the contribution for $\bm{G}=0$, it corresponds to the contribution $T_{\bm{G}= 0}^{(\eta)}$ in Eq.(\ref{PowL_5b}). With this derivation, we may write $\phi_\eta(\bm{r})$ as in Eq.(\ref{PowL_5}) with 
\begin{equation}
\label{PowL_7c}
\displaystyle T_{\bm{G}= 0}^{(\eta)} = \frac{1}{\Gamma(\frac{\eta}{2})}\frac{\pi}{A}\int_0^{\alpha^2} \frac{dt}{t^{(2-\frac{\eta}{2})}}e^{-z^2 t}
\end{equation}
and
\begin{table}
\begin{center}
\footnotesize
\begin{tabular}{|c|cc|}
\hline
\hline
&&\\
Power & Contributions depending  & Divergences\\
& on $z$ &\\
&&\\
\hline
&&\\
 $\eta > 2$ & $\displaystyle \frac{1}{\Gamma(\frac{\eta}{2})}\frac{\pi}{A}\mid z\mid^{(2-\eta)}\gamma\left(\frac{\eta}{2}-1,\alpha^2 z^2\right)$ & None\\
&&\\
$\eta=4$ & $\displaystyle \frac{\pi}{A}\frac{1}{z^2}\left(1-e^{-\alpha^2 z^2} \right) $ & None\\
&&\\
$\eta=3$ &  $\displaystyle \frac{2\pi}{A}\frac{1}{\mid z\mid} \mbox{ erf}(\alpha \mid z\mid) $& None\\
&&\\
$\eta=2$ & $\displaystyle\frac{\pi}{A}\left[\gamma+2\ln\mid z\mid - \mbox{ Ei}\left(-\alpha^2 z^2\right) \right]$ & $\displaystyle \frac{2\pi}{A}\ln\epsilon$ \\
&&\\
$\eta=1$ & $\displaystyle-\frac{2\pi}{A}\left[ \mid z\mid \mbox{ erf}(\alpha \mid z\mid)+\frac{e^{-\alpha^2 z^2}}{\alpha\sqrt{\pi}}\right]$ & $\displaystyle \frac{2\sqrt{\pi}}{A}\frac{1}{\epsilon}$ \\
&&\\
\hline
\hline
\end{tabular}
\end{center}
\caption[Contributions non triviales en $z$ des sommes d'Ewald quasi-bidimensionnelles pour $\bm{G}=0$.]{\small Non trivial contributions depending on $z$ in quasi-two dimensional Ewald summations and asymptotic behaviour for $\bm{G}=0$.}
\label{Tab3D_Powlaw_2}
\normalsize
\end{table}
\begin{equation}
\label{PowL_7d}
\begin{array}{ll}
\displaystyle\Phi_G^{(Q2)}(\eta,\alpha, z ; G)&\displaystyle  =\frac{1}{\Gamma(\frac{\eta}{2})}\frac{\pi}{A}\int_0^{\alpha^2}\frac{dt}{t^{(2-\frac{\eta}{2})}}e^{-z^2 t-\frac{G^2}{4t}}\\
&\\
&\displaystyle =\frac{1}{\Gamma(\frac{\eta}{2})}\frac{\pi}{A}\left(\frac{G^2}{4}\right)^{(\frac{\eta}{2}-1)}\int_{G^2/4\alpha^2}^\infty \frac{dt}{t^{\eta/2}}e^{-t-\frac{G^2 z^2}{4t}}
\end{array}
\end{equation}
Few analytical forms of $T_{\bm{G}= 0}^{(\eta)}$, with their dependence on $z$ and their asymptotic expension in $\epsilon$ (if $\eta \leq 2$) are given in Table \ref{Tab3D_Powlaw_2} for few integer values of $\eta$.\\ 
Two particular cases can easily be obtained with Eq.(\ref{PowL_7d}) : $z=0$, that corresponds to the Ewald method for a two dimensional system (see also \cite{Gao:97}) ; thus we have
\begin{equation}
\label{PowL_8}
\displaystyle \Phi_G^{(2)}(\eta,\alpha ; G) = \Phi_G^{(Q2)}(\eta,\alpha, 0 ; G)= \frac{1}{\Gamma(\frac{\eta}{2})}\frac{\pi}{A}\left(\frac{G}{2}\right)^{(\eta-2)}\Gamma\left(1-\frac{\eta}{2},\frac{G^2}{4\alpha^2}\right)
\end{equation}
and the limit $\alpha\rightarrow \infty$, that gives the Nijboer-de Wette representation for inverse power law interactions in quasi-two dimensional systems \cite{Nijboer:58a,Nijboer:58b} and then we found
\begin{equation}
\label{PowL_9} 
\begin{array}{ll}
\displaystyle \Phi_{\mbox{\tiny NdW},G}^{(Q2)}(\eta, z; G)&\displaystyle=\Phi_G^{(Q2)}(\eta,\infty, z ; G)\\
&\\
&\displaystyle=\frac{1}{\Gamma(\frac{\eta}{2})}\frac{2\pi^{1/2}}{A}\left(\frac{G}{2}\right)^{(\frac{\eta}{2}-1)}\mid z\mid^{(1-\frac{\eta}{2})}\mbox{ K}_{(\frac{\eta}{2}-1)}\left(G\mid z\mid\right)
\end{array}
\end{equation}
The general case is more complicated to obtain, it is necessary to use incomplete Bessel or generalized incomplete gamma functions \cite{Harris:08,Chaudhry:94,Chaudhry:96}. Following the definition given by Harris in ref.\cite{Harris:08}, the incomplete Bessel function is defined as
\begin{equation}
\label{PowL_10}
\displaystyle \mbox{ K}_{\nu}(x,y)=\int_{1}^{\infty}\frac{dt}{t^{\nu+1}}e^{-xt-\frac{y}{t}}
\end{equation}
Then, with the second integral in Eqs.(\ref{PowL_7d}), we find
\begin{equation}
\label{PowL_11}
\displaystyle\Phi_G^{(Q2)}(\eta,\alpha, z ; G)=\frac{\alpha^{(\eta-2)}}{\Gamma(\frac{\eta}{2})}\frac{\pi}{A}\mbox{ K}_{(\frac{\eta}{2}-1)}\left(\frac{G^2}{4\alpha^2},\alpha^2 z^2\right)
\end{equation}
According to the connection between the generalized incomplete gamma function and the incomplete Bessel function, defined as (Eq.(6) of ref.\cite{Harris:08})
\begin{equation}
\label{PowL_12}
\displaystyle \mbox{ K}_{\nu}(x,y)=x^{\nu}\mbox{ }\Gamma\left(-\nu, x ; xy \right)
\end{equation}
Eq.(\ref{PowL_11}) can also be written as 
\begin{equation}
\label{PowL_13}
\displaystyle\Phi_G^{(Q2)}(\eta,\alpha, z ; G)=\frac{2^{(2-\eta)}}{\Gamma(\frac{\eta}{2})}\frac{\pi}{A}\frac{1}{G^{(2-\eta)}}\mbox{ }\Gamma\left(1-\frac{\eta}{2},\frac{G^2}{4\alpha^2} ; \frac{G^2 z^2}{4} \right)
\end{equation}
If $\eta=1$, the inverse power law interaction corresponds to the Coulomb interaction ; one can easily verify that Eqs.(\ref{PowL_11}) and (\ref{PowL_13}) allow to recover the Ewald sums for Coulomb interactions in quasi-two dimensional systems \cite{Parry:75,Smith:88,Harris:98,Grybowski:00,Mazars:05b} (see also Theorem 7 in ref.\cite{Chaudhry:94}).\\
On Table \ref{Tab3D_Powlaw_3}, we give some analytical formulas for the computation of the reciprocal space contributions in quasi-two dimensional  Ewald summations for inverse power law interactions. For numerical implementations, convenient algorithms for the computations of the incomplete Bessel functions are needed (see for instance refs.\cite{Harris:09,Fripiat:09}).\\
\begin{table}
\begin{center}
\tiny
\begin{tabular}{|c|c|c|}
\hline
\hline
&&\\
Power & $\Phi_G^{(Q2)}(\eta,\alpha, z ; G)$   & References\\
&&\\
\hline
\hline
&&\\
$\eta$ & $\displaystyle \frac{\pi}{A}\frac{\alpha^{(\eta-2)}}{\Gamma(\frac{\eta}{2})}\mbox{ K}_{(\frac{\eta}{2}-1)}\left(\frac{G^2}{4\alpha^2},\alpha^2 z^2\right)$ & Eq.(\ref{PowL_11}) \\ 
&&\\
            & $\displaystyle \frac{\pi}{A}\frac{2^{(2-\eta)}}{\Gamma(\frac{\eta}{2})}\frac{1}{G^{(2-\eta)}}\mbox{ }\Gamma\left(1-\frac{\eta}{2},\frac{G^2}{4\alpha^2} ; \frac{G^2 z^2}{4} \right)$ & Eq.(\ref{PowL_13})\\ 
&&\\
\hline
\hline
&&\\
$\eta=4$ & $\displaystyle \frac{\pi}{A}\left[\frac{G}{2\mid z\mid}\mbox{ K}_{1}\left(G \mid z\mid\right) \right.$& Eq.(20) in ref.\cite{Harris:08} \\ 
&$\displaystyle \left. -\frac{\alpha^2 e^{-(Y^2+X^2)}}{2 Y^2} \left[1+\sum_{m=0}^{\infty}(X^2+Y^2-m)\frac{(X^2-Y^2)^m}{m!}\mbox{ Q}_m\left(\frac{X^2+Y^2}{X^2-Y^2}\right) \right]\right]$ & (Q$_m(z)$ is a Legendre function (cf. section 8.7 of ref.\cite{Gradshteyn:book:00})\\
&&\\
      & with $X=G/2\alpha$ and $Y=\alpha z$ & \\
&&\\
\hline
&&\\
$\eta=3$ &  $\displaystyle \frac{\pi}{A} \frac{1}{\mid z\mid}\left[ e^{-Gz}\mbox{ erfc}\left(\frac{G}{2\alpha}-\alpha z\right)-e^{Gz}\mbox{ erfc}\left(\frac{G}{2\alpha}+\alpha z\right) \right]$ & Eq.(28) in ref.\cite{Chaudhry:94} \\ 
 & & and also in Ewald methods for dipolar interactions.\\
&&\\
\hline
&&\\
$\eta=2$ &$\displaystyle\frac{\pi}{A}\left[\mbox{ K}_{0}\left(G \mid z\mid\right)- e^{-(Y^2+X^2)}\sum_{m=0}^{\infty}\frac{(X^2-Y^2)^m}{m!}\mbox{ Q}_m\left(\frac{X^2+Y^2}{X^2-Y^2}\right)\right] $ & Eq.(19) in ref.\cite{Harris:08}\\ 
&&\\
                 & with $X=G/2\alpha$ and $Y=\alpha z$ & \\
&&\\
\hline
&&\\
$\eta=1$ & $\displaystyle \frac{\pi}{A} \frac{1}{G}\left[ e^{-Gz}\mbox{ erfc}\left(\frac{G}{2\alpha}-\alpha z\right)+e^{Gz}\mbox{ erfc}\left(\frac{G}{2\alpha}+\alpha z\right) \right]$ & Theorem 7 in ref.\cite{Chaudhry:94} \\
&& and Ewald methods for Coulomb interactions.  \\ 
&&\\
\hline
\end{tabular}
\end{center}
\caption[Contributions Espace reciproque quasi-2D.]{\small Analytical formulas for the reciprocal space contributions in quasi-two dimensional  Ewald summations for inverse power law interactions $1/r^\eta$ for some values of $\eta$. The column 'References' gives some sources useful for analytical computations of $\Phi_G^{(Q2)}(\eta,\alpha, z ; G)$. As shown with Eq.(\ref{PowL_5}), the real space contribution to Ewald summations is given in Table \ref{Tab3D_Powlaw_1}.}
\label{Tab3D_Powlaw_3}
\normalsize
\end{table}
To close this section, we want to outline that the equivalence between both derivations allows to obtain a new analytical relation for the Fourier transform of incomplete gamma function. From Eq.(\ref{PowL_7}), we find
\begin{equation}
\label{PowL_14}
\hspace{-1.in}\begin{array}{ll}
\displaystyle \frac{1}{2\pi}\int_{-\infty}^{+\infty}dk \frac{\Gamma\left(\frac{3-\eta}{2},\frac{1}{4\alpha^2}(G^2+k^2)\right)}{(G^2+k^2)^{(3-\eta)/2}}e^{j k z} &\displaystyle  = \frac{(2\alpha)^{(\eta-2)}}{2\sqrt{\pi}}\mbox{ K}_{(\frac{\eta}{2}-1)}\left(\frac{G^2}{4\alpha^2},\alpha^2 z^2\right)\\
&\\
&\displaystyle = \frac{1}{2\sqrt{\pi}}\frac{1}{G^{(2-\eta)}}\Gamma\left(1-\frac{\eta}{2},\frac{G^2}{4\alpha^2} ; \frac{G^2 z^2}{4} \right)
\end{array}
\end{equation}
If one takes $\alpha=1$ and $G=0$, we recover the Fourier transform of the incomplete gamma function for $\nu>1/2$, as
\begin{equation}
\label{PowL_15}
\displaystyle \frac{1}{2\pi}\int_{-\infty}^{+\infty}dk \mbox{ }\frac{\Gamma\left(\nu,\frac{k^2}{4}\right)}{k^{2\nu}}\mbox{ }e^{j k z}= \frac{1}{\mid z\mid \sqrt{\pi}}\left(\frac{z}{2}\right)^{2\nu}\gamma\left(\nu-\frac{1}{2},z^2\right)
\end{equation}
where we have used the definition of the incomplete gamma function $\gamma(\mu,y)$
\begin{equation}
\label{PowL_16}
\displaystyle \gamma(\mu,y)=\int_0^y dx\mbox{ }x^{(\mu-1)}\mbox{ }e^{-x}=y^{\mu}\mbox{ K}_{\mu}(0,y)
\end{equation}
Conversely, an analytical demonstration of Eq.(\ref{PowL_14}) based on the properties of incomplete and generalized incomplete gamma functions will demonstrate the equivalence of both derivations for inverse power law interactions (already shown for Coulomb, Yukawa and Dipolar interactions \cite{Mazars:07d,Mazars:07a,Weis:book:05c}). Such a pure analytical derivation of Eq.(\ref{PowL_14}) has not yet been achieved \cite{Chaudhry:02}.\\
The equivalence between both derivations may also be justified directly by the relation between Poisson-Jacobi identities in three and two dimensions. For one dimensional periodicity, the Poisson-Jacobi is \cite{Mazars:01b}  
\begin{equation}
\label{Poisson_1D}
\sum_{n=-\infty}^{+\infty}e^{-\mid z+ n L_z\mid^2 t}=\frac{1}{L_z}\left(\sqrt{\frac{\pi}{t}}\right)\sum_{k_z}e^{j k_z z}\exp\left(-\frac{k_z^2}{4}\frac{1}{t}\right)
\end{equation}
then, multiplying Eq.(\ref{PoissonJacobi2D}) with Eq.(\ref{Poisson_1D}), we obtain the Poisson-Jacobi for three dimensions Eq.(\ref{PoissonJacobi3D}). This agrees with the separations done in Eqs.(\ref{Para_2D},\ref{Decompo_2D}).

\section{Electroneutrality and IR-divergences.}

The IR-divergences, that occur in Ewald sums of inverse power-law interactions when $\eta\leq d$, with $d$ the dimension of the periodicity, are related to the conditional convergence properties of the lattice sums Eq.(\ref{PowL_1}). These divergences are an artefact due to the approximation of the finite lattice sum by an infinite lattice sum (Ewald sum) \cite{Smith:08,Herce:07}. For coulomb interactions these divergences in the infinite lattice sums are cancelled in the computation of energy and forces when the system in the basic cell is electroneutral. Two classical models are used with coulomb interactions : the restricted primitive model of electrolytes (RPM) and the one component plasma model (OCP) \cite{Baus:80,Totsuji:78,Weis:98}. The RPM is made of $2N_0$ hard spheres that carry electric point charge at the centre of the sphere, $N_0$ particles carry a charge $+Q$ and the other $N_0$ particles carry a charge $-Q$. We define the $\eta$-RPM similarly to the standard RPM, but with interaction between charges as inverse power law interactions. In the OCP model, there are $N$ point particles carrying all the same charge $Q$ and a constant volume density of charge $\rho_0$. In the $\eta$-OCP model, the system is defined as the standard OCP model and interactions between point particles are inverse power law interactions.\\  
In the following, we first compute the energy of a system of $N$ point particles carrying pseudo-charges $Q_i$ in a system with periodic boundary conditions in all three directions of the space. Then, in subsection 4.1, we make the same computation but for a quasi-two dimensional system where electroneutrality is fulfilled with the help of a constant planar surface density of pseudo-charge. In subsection 4.2, we apply the analytical results of section 3. to a monolayer system made of $N$ point particles carrying the same pseudo-charge $Q$, this model is a generalization of the two dimensional One-Component plasma ($\eta$-OCP). Few computations using Monte-Carlo simulations for monolayers of $\eta$-RPM and $\eta$-OCP systems are reported in subsection 4.2.\\
Let the interaction energy between two pseudo-charges be
\begin{equation} 
E_{ij}=Q_iQ_j \phi_\eta\left(\frac{\bm{r}_{ij}}{r_0}\right)
\end{equation}
where $r_0$ is a typical length defined by the geometry of particles, for instance the diameter of hard spheres in the $\eta$-RPM model. In the following, all length are measured in unit of $r_0$ (i.e. we set $r_0=1$). The energy of the system with tridimensional periodic boundary conditions is given by 
\begin{equation}
\label{U3D}
\displaystyle E_{\eta}=\frac{1}{2} \sum_{i=1}^N \sum_{j\neq i}\mbox{}^{\prime} Q_i Q_j\phi_\eta(\bm{r}_{ij})+\frac{1}{2}\sum_{i=1}^NQ_i^2 \phi_{\eta}^{(0)}
\end{equation}
with $\phi_{\eta}^{(0)}$ the contribution of the interaction of a charge with its own periodic images. This quantity is defined and computed as for $\phi_\eta(\bm{r})$, thus we find
\begin{equation}
\label{self_3D}
\hspace{-0.5in}\begin{array}{ll}
\displaystyle \phi_\eta^{(0)} =\sum_{\bm{L}_{\bm{n}}\neq 0}\frac{1}{\mid \bm{L}_{\bm{n}}\mid^\eta}&\displaystyle=\sum_{\bm{L}_{\bm{n}}} \Phi_R^{(3)}\left(\eta,\alpha ; \mid \bm{L}_{\bm{n}}\mid\right)+\sum_{\bm{k}\neq 0}\Phi_k^{(3)}(\eta,\alpha ; k)\\
&\\
&\displaystyle -\frac{2}{\eta\Gamma\left(\frac{\eta}{2}\right)}\alpha^{\eta}+\frac{1}{\Gamma\left(\frac{\eta}{2}\right)}\frac{\pi^{3/2}}{V}\int_0^{\alpha^2}\frac{dt}{t^{(5-\eta)/2}}
\end{array}
\end{equation}
The last contribution in the previous equation has the same IR-divergence than in Eq.(\ref{PowL_4}). When it is grouped with the other contributions of Eq.(\ref{U3D}), we have 
\begin{equation}
\label{electroN}
\frac{1}{\Gamma\left(\frac{\eta}{2}\right)}\frac{\pi^{3/2}}{V}\left(\sum_{i=1}^N Q_i \right)^2\int_0^{\alpha^2}\frac{dt}{t^{(5-\eta)/2}}\equiv 0
\end{equation}
that is cancelled if $\sum_{i=1}^N Q_i =0$. It would be more rigorous to consider that this equation is an indeterminate of  the form $0\times\infty$ ; it corresponds to the property of conditional convergence of lattice sums for inverse power-law interactions, this is related to the macroscopic boundary condition and to the particularity of the systems (see for instance ref.\cite{Herce:07,Smith:08} for Coulomb interaction, $\eta=1$).\\
Then the energy can be written as 
\begin{equation}
\label{U3D_ex}
\begin{array}{ll}
\displaystyle E_{\eta}&\displaystyle=\frac{1}{2} \sum_{j=1}^N\sum_{i=1}^N  \sum_{\mbox{\small $\bm{L}_{\bm{n}}$} }\mbox{}^{\prime} Q_i Q_j\Phi_R^{(3)}\left(\eta,\alpha ; \mid\bm{r}_{ij}+\bm{L}_{\bm{n}}\mid\right)\\
&\\ 
&\displaystyle +\sum_{\bm{k}\neq 0} \Phi_k^{(3)}(\eta,\alpha ; k) \left|\sum_{i=1}^N Q_i\exp\left(j\bm{k}.\bm{r}_i\right)\right|^2- \frac{\alpha^{\eta}}{\eta\Gamma\left(\frac{\eta}{2}\right)} \left(\sum_{i=1}^{N}Q_i^2\right)
\end{array}
\end{equation}
where the prime in the first contribution indicates that the term $i=j$ is not included when $\bm{L}_{\bm{n}}\equiv\bm{L}_{\bm{0}}\equiv\bm{0}$. We have also made a factoring of the reciprocal space contributions as 1-particle summations, this is an important property for the numerical efficiency of the summations as it allows to the computational time needed for obtaining the reciprocal contribution to the energy to scale as $N$ and not as $n^2$. One may verify easily that for $\eta=1$, we recover the Ewald summations for Coulomb interactions, Eq.(\ref{U3D_ex}) allows to obtain the energy of the $\eta$-RPM tridimensional systems.\\
In the $\eta$-One Component Plasma model ($\eta$-OCP) all point-particles carry the same charge $Q$ and electroneutrality is achieved by the constant volume charge density $\rho_0$ such as 
\begin{equation}
NQ+\rho_0 V = 0
\end{equation}
For this system, one may show easily that the IR-divergence is cancelled as in Eq.(\ref{electroN}), written as
\begin{equation}
\label{electroN_OCP}
\displaystyle\frac{1}{\Gamma\left(\frac{\eta}{2}\right)}\frac{\pi^{3/2}}{V}\left(N Q +\rho_0 V \right)^2\int_0^{\alpha^2}\frac{dt}{t^{(5-\eta)/2}}\equiv 0
\end{equation}
The energy $E(\mbox{$\eta$-OCP})$ is given by using Eq.(\ref{U3D_ex}) with $Q_i=Q$, for $0<\eta<3$, we obtain 
\begin{equation}
\label{E_eta_OCP}
\displaystyle E(\mbox{$\eta$-OCP}) = E_{\eta} -\frac{\pi^{3/2}\alpha^{(\eta-3)}}{(3-\eta)\Gamma(\frac{\eta}{2})}\frac{N^2 Q^2}{V}
\end{equation}
and for $\eta=3$, we have
\begin{equation}
\label{E_3_OCP}
\displaystyle E(\mbox{3-OCP}) = E_{3} -2\pi\frac{N^2 Q^2}{V}\ln\alpha^2
\end{equation}
In Eqs.(\ref{E_eta_OCP},\ref{E_3_OCP}), the contributions that scale as $N^2$ stem from the interaction of particles with the neutralizing background. For $\eta>3$, no IR-divergence is present and the energy of the system can be obtained from Eq.(\ref{U3D_ex}).

\subsection{Electroneutrality in quasi-two dimensional systems.}

To extend the tridimensional $\eta$-OCP model to quasi-two dimensional systems, one has to choose a particular neutralizing background to fulfill electroneutrality ; a constant volume charge density $\rho_0$ is inconsistent with the symmetry of quasi-two dimensional systems. Several choices can be done \cite{Mazars:07a} ; in the following, we choose the neutralizing background as a constant planar surface density $\sigma_0$ of pseudo-charges. With this choice, the electroneutrality of the system reads as
\begin{equation}
NQ+\sigma_0 A = 0
\end{equation}
and the energy of the $\eta$-OCP, for $0<\eta\leq 2$, can be computed as 
\begin{equation}
\label{E_eta_OCP_int}
\hspace{-1.in}\begin{array}{ll}
\displaystyle E(\mbox{$\eta$-OCP})&\displaystyle =\frac{Q^2}{2}\sum_{i=1}^{N}\sum_{j=1}^{N}\sum_{\mbox{\small $\bm{S}_{\bm{n}}$}}\mbox{}^{\prime}\frac{1}{\mid\bm{r}_{ij}+\bm{S}_{\bm{n}}\mid^\eta}+Q\sigma_0\sum_{i=1}^{N}\int_{\mbox{\small $\bm{S}_{\bm{0}}$}} d\bm{s} \sum_{\mbox{\small $\bm{S}_{\bm{n}}$}}\frac{1}{\mid\bm{r}_{i}-\bm{s}+\bm{S}_{\bm{n}}\mid^\eta}\\
&\\
&\displaystyle + \frac{\sigma_0^2}{2}\int_{\mbox{\small $\bm{S}_{\bm{0}}$}} d\bm{s}\int_{\mbox{\small $\bm{S}_{\bm{0}}$}} d\bm{s}'\sum_{\mbox{\small $\bm{S}_{\bm{n}}$}}\frac{1}{\mid\bm{s}'-\bm{s}+\bm{S}_{\bm{n}}\mid^\eta}
\end{array}
\end{equation}
To compute the lattice sums, we apply the methods of section 2 and 3 ; with the notations of Eq.(\ref{Para_2D}), for $0<\eta<2$, we find
\begin{equation}
\label{E_eta_OCP_Q2D}
\hspace{-1.in}\begin{array}{ll}
\displaystyle E(\mbox{$\eta$-OCP})&\displaystyle =\frac{Q^2}{2}\sum_{i=1}^{N}\sum_{j=1}^{N}\sum_{\mbox{\small $\bm{S}_{\bm{n}}$}}\mbox{}^{\prime}\Phi_R^{(3)}\left(\eta,\alpha ; \mid\bm{r}_{ij}+\bm{S}_{\bm{n}}\mid\right)-\left(\frac{\alpha^{\eta}}{\eta\Gamma\left(\frac{\eta}{2}\right)}+\frac{\pi}{A}\frac{\alpha^{(\eta-2)}}{(2-\eta)\Gamma\left(\frac{\eta}{2}\right)}\right)NQ^2\\
&\\
&\displaystyle+\frac{Q^2}{2}\sum_{\bm{G}\neq 0}\sum_{i=1}^{N}\sum_{j=1}^{N}e^{j\bm{k}.\bm{s}_{ij}}\Phi_G^{(Q2)}(\eta,\alpha, z_{ij} ; G)-2\frac{\pi}{A}\frac{NQ^2}{(\eta-2)}\sum_{i=1}^{N}\mid z_i \mid^{(2-\eta)}\\
&\\
&\displaystyle +\frac{\pi}{A}\frac{\alpha^{(\eta-2)}Q^2}{(2-\eta)\Gamma\left(\frac{\eta}{2}\right)}\sum_{i=1}^{N}\sum_{j\neq i}\left[e^{-\alpha^2 z_{ij}^2}+(\alpha\mid z_{ij}\mid)^{(2-\eta)}\gamma\left(\frac{\eta}{2},\alpha^2 z_{ij}^2\right)\right]
\end{array}
\end{equation}
and for $\eta=2$, we have
\begin{equation}
\label{E_2_OCP_Q2D}
\hspace{-1.in}\begin{array}{ll}
\displaystyle E(\mbox{2-OCP})&\displaystyle =\frac{Q^2}{2}\sum_{i=1}^{N}\sum_{j=1}^{N}\sum_{\mbox{\small $\bm{S}_{\bm{n}}$}}\mbox{}^{\prime}\Phi_R^{(3)}\left(2,\alpha ; \mid\bm{r}_{ij}+\bm{S}_{\bm{n}}\mid\right)-\frac{\pi NQ^2}{2A}\left(C+2\ln\alpha -\frac{\alpha^2 A}{\pi}\right)\\
&\\
&\displaystyle +\frac{Q^2}{2}\sum_{\bm{G}\neq 0}\sum_{i=1}^{N}\sum_{j=1}^{N}e^{j\bm{k}.\bm{s}_{ij}}\Phi_G^{(Q2)}(2,\alpha, z_{ij} ; G)-\frac{\pi CN^2 Q^2}{2A}\\
&\\
&\displaystyle -\frac{\pi Q^2}{2A}\sum_{i=1}^{N}\sum_{j\neq i}\mbox{ E}_{1}\left(\alpha^2 z_{ij}^2\right)-\frac{\pi Q^2}{2A}\sum_{i=1}^{N}\sum_{j\neq i}\ln z_{ij}^2-\frac{\pi Q^2}{A}\sum_{i=1}^{N}\ln z_i^2
\end{array}
\end{equation}
with $C$ the Euler's constant and $\mbox{E}_{1}(x)$ the exponential integral.\\
It is worthwhile to note that for quasi-two dimensional systems the factoring of the reciprocal part of the energy into 1-particle summations cannot be achieved because of the complicated dependence on $z$ in $\Phi_G^{(Q2)}(2,\alpha, z ; G)$.

\subsection{Monolayers and few numerical results.}

When all particles are confined in a plane, the system is a monolayer and one can factor the reciprocal part of the energy into 1-particle summations. For a system of $N$ point particles carrying a pseudo-charge $Q_i$, satisfying to the electroneutrality, one finds easily
\begin{equation}
\label{E_eta_RPM_2D}
\hspace{-1.in}\begin{array}{ll}
\displaystyle E_{\eta}^{(cc)} &\displaystyle =\frac{1}{2}\sum_{i=1}^{N}\sum_{j=1}^{N}Q_i Q_j\sum_{\mbox{\small $\bm{S}_{\bm{n}}$}}\mbox{}^{\prime}\Phi_R^{(3)}\left(\eta,\alpha ; \mid\bm{s}_{ij}+\bm{S}_{\bm{n}}\mid\right)-\frac{\alpha^{\eta}}{\eta\Gamma\left(\frac{\eta}{2}\right)}\sum_{i=1}^NQ_i^2\\
&\\
&\displaystyle +\sum_{\bm{G}\neq 0} \Phi_G^{(2)}(\eta,\alpha ; G) \left|\sum_{i=1}^N Q_i\exp\left(j\bm{G}.\bm{s}_i\right)\right|^2
\end{array}
\end{equation}
For the $\eta$-One Component Plasma model with all point particles confined in the plan with the constant surface charge density, the energy is given by 
\begin{equation}
\label{E_eta_OCP_2D}
\hspace{-0.8in}\begin{array}{ll}
\displaystyle E_{\eta} &\displaystyle =\frac{Q^2}{2}\sum_{i=1}^{N}\sum_{j=1}^{N}\sum_{\mbox{\small $\bm{S}_{\bm{n}}$}}\mbox{}^{\prime}\Phi_R^{(3)}\left(\eta,\alpha ; \mid\bm{s}_{ij}+\bm{S}_{\bm{n}}\mid\right)-\frac{\alpha^{\eta}}{\eta\Gamma\left(\frac{\eta}{2}\right)}NQ^2\\
&\\
&\displaystyle +Q^2\sum_{\bm{G}\neq 0} \Phi_G^{(2)}(\eta,\alpha ; G) \left|\sum_{i=1}^N \exp\left(j\bm{G}.\bm{s}_i\right)\right|^2-\frac{\pi}{A}\frac{\alpha^{(\eta-2)}}{(2-\eta)\Gamma\left(\frac{\eta}{2}\right)}N^2Q^2
\end{array}
\end{equation}
\begin{table}
\begin{center}
\footnotesize
\begin{tabular}{|ll|cll|}
\hline
\hline
&&&&\\
& $\eta$-RPM & & &  $\eta$-OCP \\
&&&&\\
$\eta$ & $\beta U/N$ & $\epsilon^{\dag}$ & $\eta$ & $\beta U/N$ \\
&&&&\\
\hline
&&&&\\
0.25         &  -10.47(3) & $5\times 10^2$ & 0.5 & -10.76(2)\\
0.5           &  -11.36(3) & $5\times 10^2$ & 1.0$^{\ddag}$ & -16.67(3) \\
0.75         &  -12.29(3) & $5\times 10^2$ & 1.5 & -33.04(3)\\
1.0$^{*}$ &  -13.23(3)  &                          &       & \\
1.25         & -14.21(5)  & $5\times 10^3$ & 0.5 & -35.26(3) \\
1.5           & -15.2(2)  & $5\times 10^3$ & 1.0$^{\ddag}$ & -54.18(3) \\
1.75         & -16.3(3)  & $5\times 10^3$ & 1.5 & -105.97(3)\\
&&&&\\
\hline
\hline
\end{tabular}\\[0.07in]
\footnotesize
 $^{\dag}$  In this table, $\epsilon$ is defined as in ref.\cite{Totsuji:78} : $\epsilon=2\pi\rho Q^4/T^2$. \\
 $^{*}$ cf. ref.\cite{Weis:98}, Table II, Line 15.\\
 $^{\ddag}$ cf. ref.\cite{Totsuji:78}, Table II.
\end{center}
\caption{\small  Average energies for $\eta$-RPM and $\eta$-OCP models in Monte Carlo computations. The numbers in brackets give the accuracy on the last digit of the averages. For the $\eta$-RPM, the average energy is defined by $U=<E_{\eta}^{(cc)}>_{MC}$ with $E_{\eta}^{(cc)}$ given by Eq.(\ref{E_eta_RPM_2D}) ; for the computations reported in this table, one has : $N=2N_0=1024$, $\rho=2N_0/A = 0.6$, $Q^2=20$ (it corresponds to $\beta^*=20$ in notations of ref.\cite{Weis:98}). For the $\eta$-OCP model, the average energy is defined as $U=<E_\eta>_{MC}$ with Eq.(\ref{E_eta_OCP_2D}) ; one has : $N=1024$, $\pi\rho=\pi N/A =1.0$, $Q\simeq 3.98$ for $\epsilon=5\times 10^2$ and $Q\simeq7.071$ for $\epsilon=5\times 10^3$.}
\label{TabMC_Powlaw_4}
\normalsize
\end{table}
\begin{figure}
\begin{tabular}{ll}
\hspace{-1.5in}\centerline{(a)\includegraphics[width=2.6in]{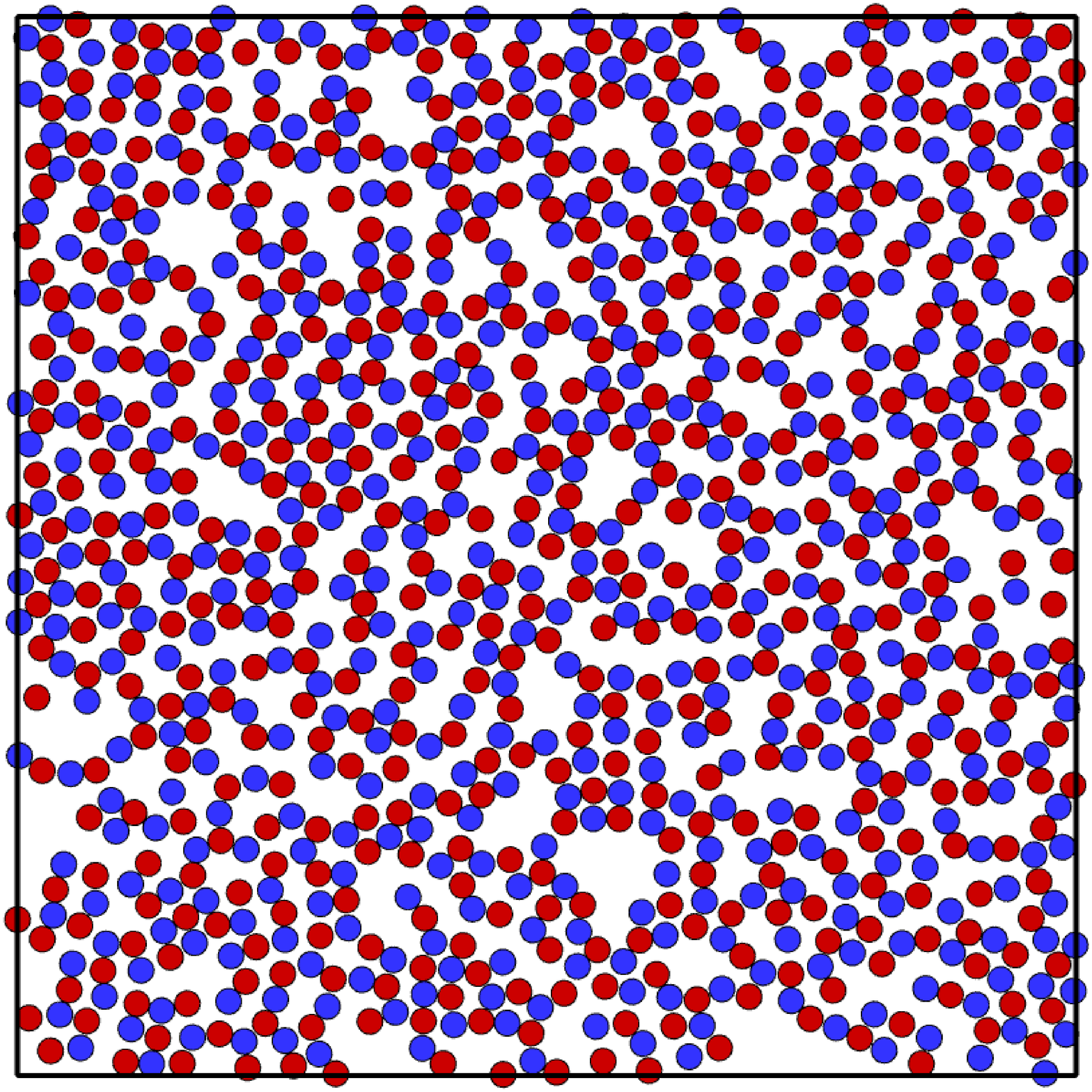}}&\hspace{-3.in}\centerline{(b)\includegraphics[width=2.6in]{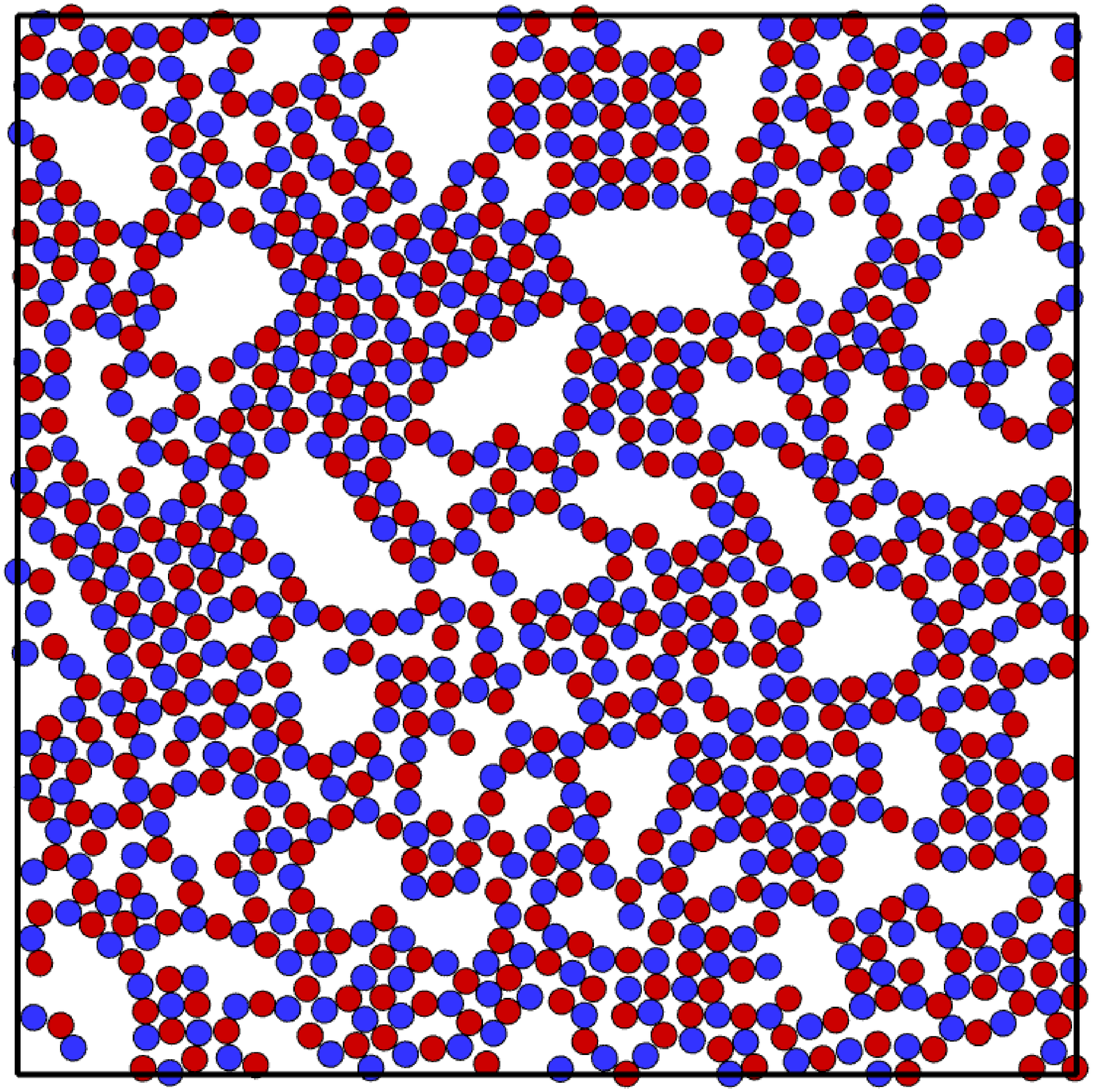}}\\
\hspace{-1.5in}\centerline{(c)\includegraphics[width=3.55in]{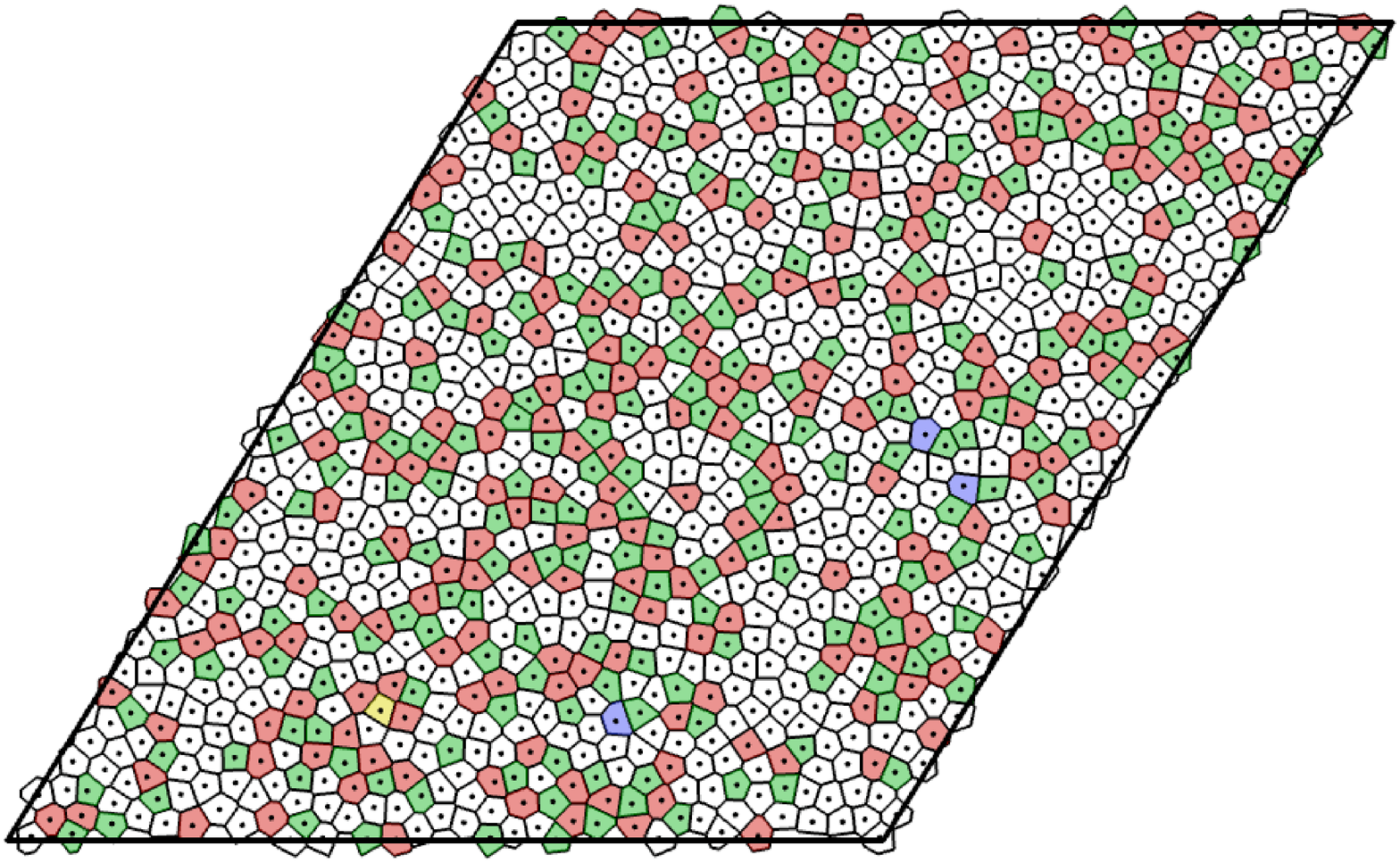}}&\hspace{-3.1in}\centerline{(d)\includegraphics[width=3.in]{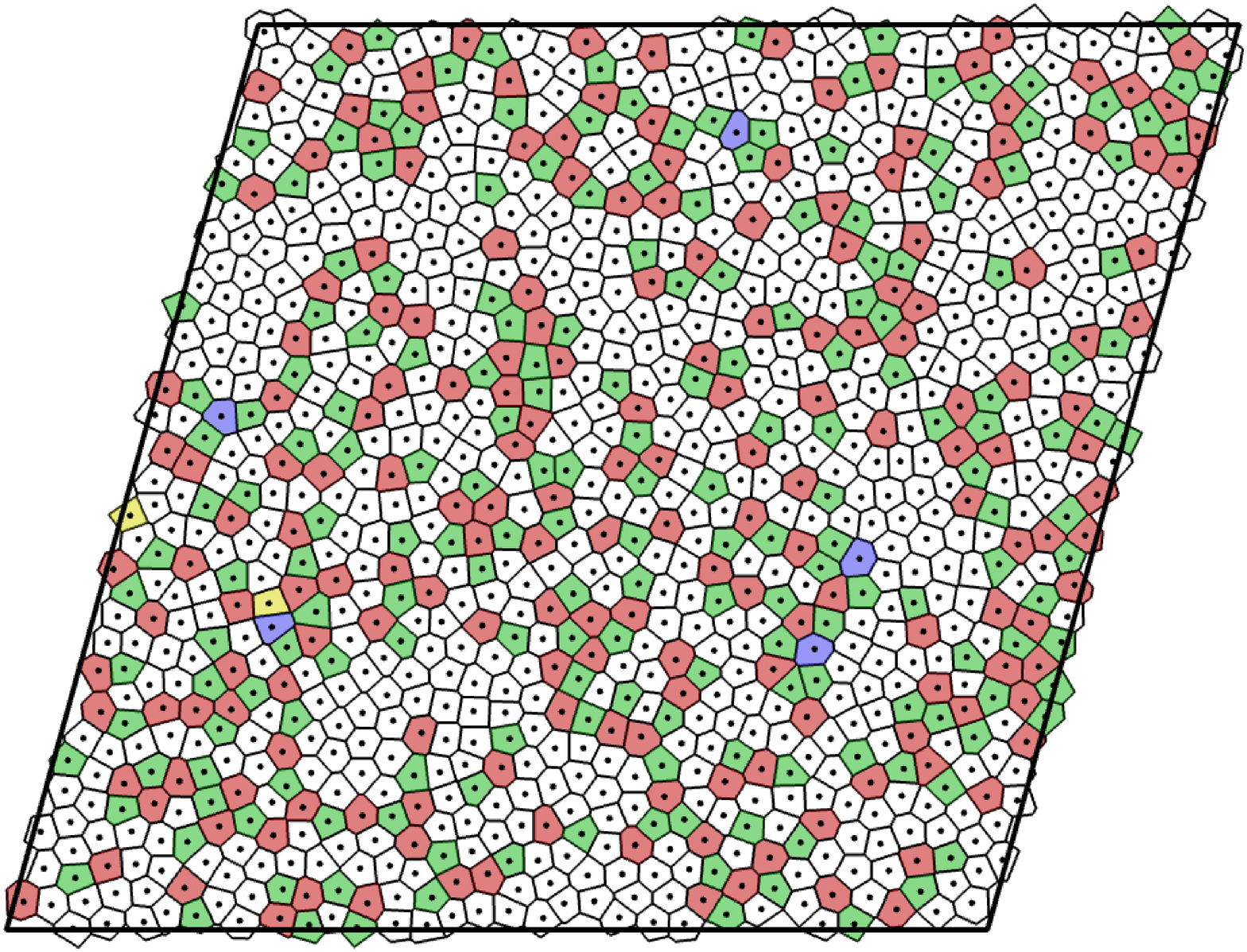}}\\
\end{tabular}
\caption{\small Snapshots of  $\eta$-RPM and $\eta$-OCP monolayer models. The sides of the simulation box are represented by thick black lines, periodic boundary conditions are applied. (a-b) : $\eta$-RPM, blue disks are particles carrying a negative charge ($-Q$) and red disks a positive charge ($+Q$). For both snapshots, $N=2N_0=1024$, $\rho=2N_0/A = 0.6$, $Q^2=20$ (cf. Table \ref{TabMC_Powlaw_4}) ; (a) $\eta=1.0$ and (b) $\eta=1.75$. (c-d) : $\eta$-OCP model with Voronoi construction, for both snapshots : $N=1024$, $\pi\rho=\pi N/A =1.0$ and $\epsilon=5\times 10^3$ (cf. Table \ref{TabMC_Powlaw_4}). Voronoi cells with four sides are represented in yellow, those with five sides are represented in green, those with six sides in white, those with seven sides in red and those with eight sides in blue. Point particles are represented by a small dot. (c) $\eta=1.0$ and (d) $\eta=1.5$. }
\label{Fig_PowLaw_1}
\end{figure}
\begin{figure}
\includegraphics[width=6.5in]{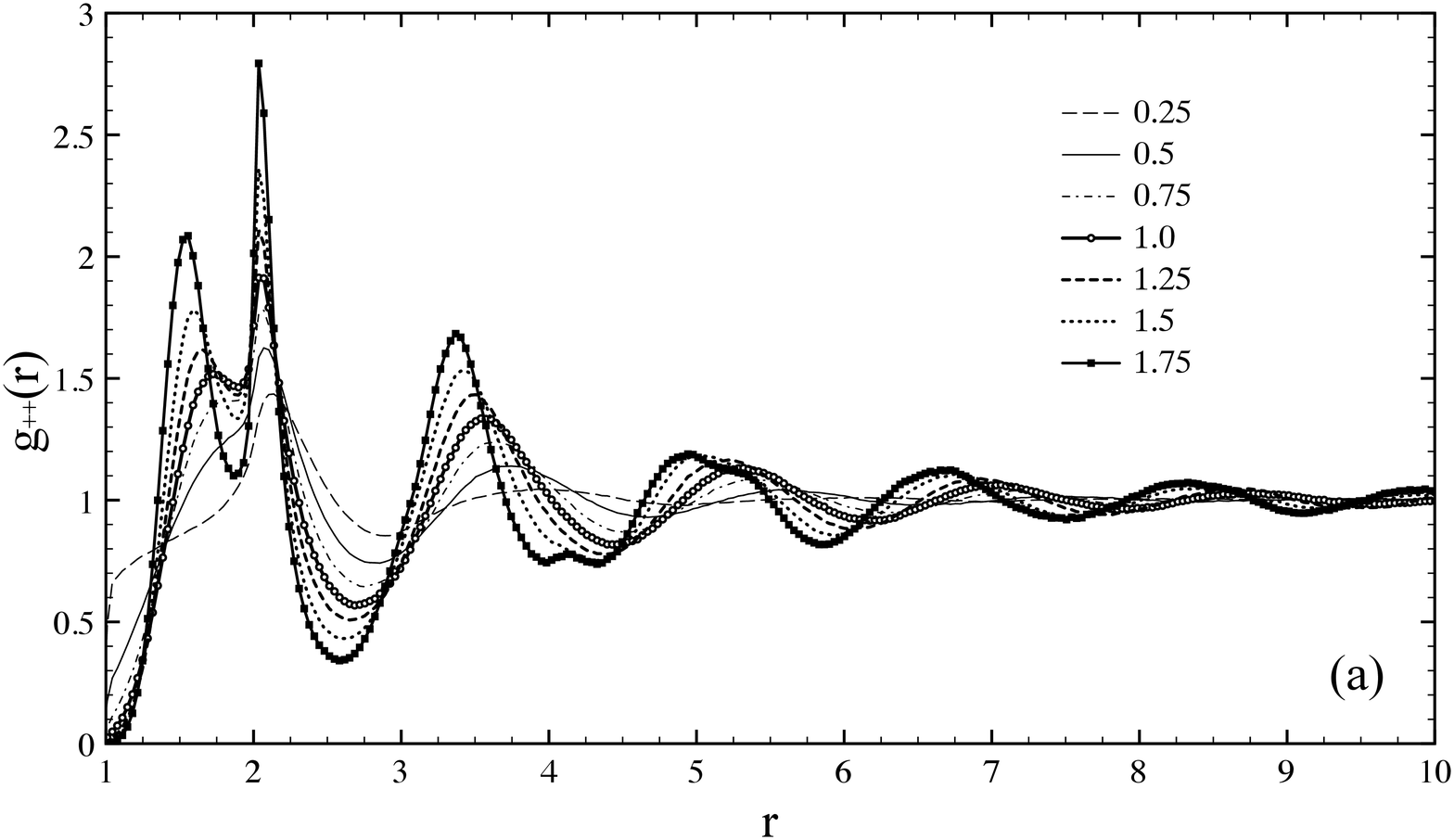}\\
\includegraphics[width=6.5in]{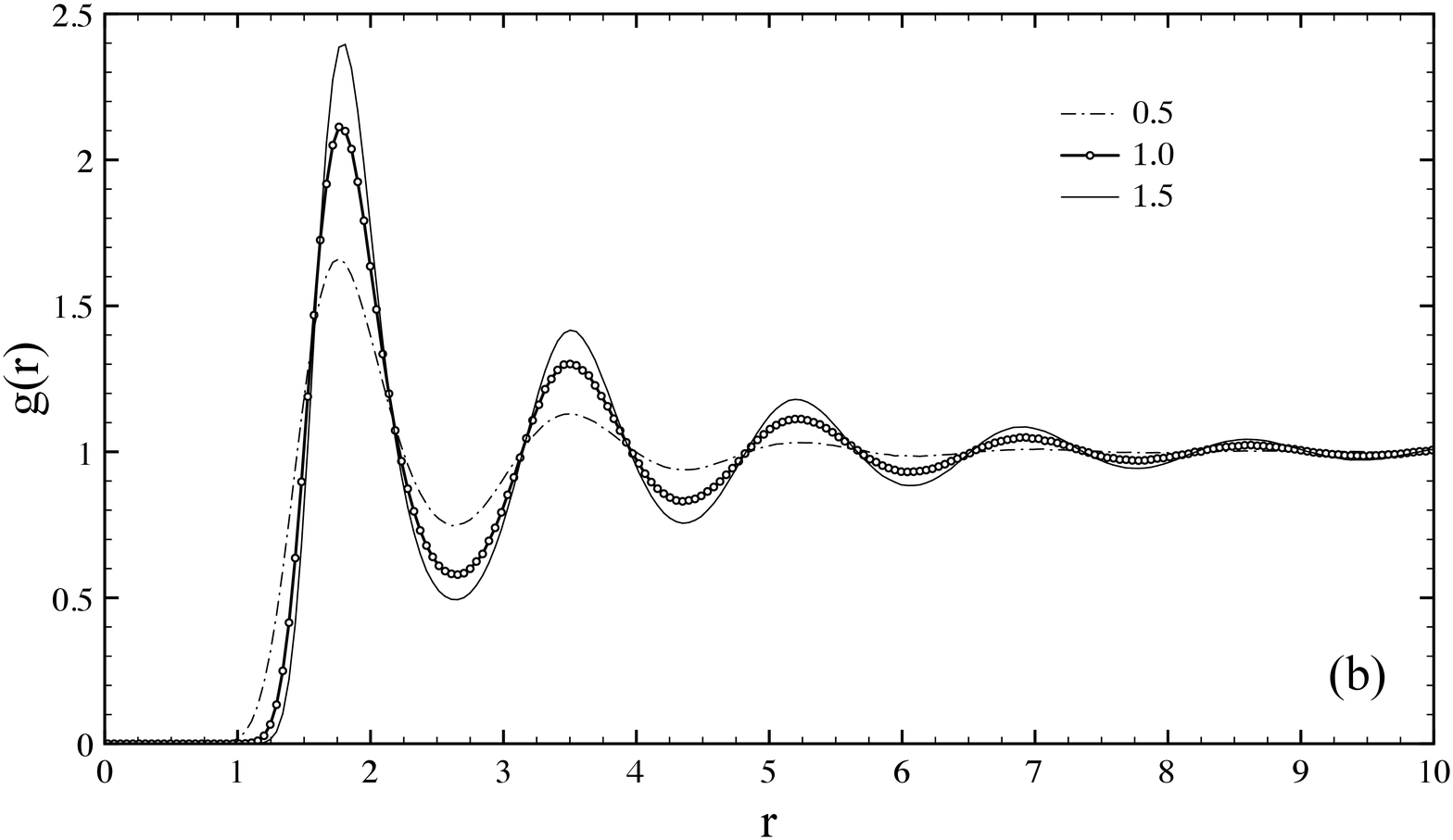}
\caption{\small  Pair correlation functions for $\eta$-RPM and $\eta$-OCP models. The thermodynamic states of each system are the same as those reported on Table \ref{TabMC_Powlaw_4}. For each system, the values of $\eta$ are given in legends. (a) Pair correlation functions $g_{++}(r)$ between positive ions in the $\eta$-RPM model ; for $\eta=1.0$, $g_{++}(r)$ can be compared with the standard RPM model (see FIG.10(a) of ref.\cite{Weis:98}). (b) Pair correlation functions $g(r)$ between point particles in the $\eta$-OCP model with $\epsilon=2\pi\rho Q^4/T^2= 5\times 10^3$ ;  for $\eta=1.0$, $g(r)$ is the same as the one given in the FIG.2 of ref.\cite{Totsuji:78}.}
\label{Fig_PowLaw_2}
\end{figure}
As outlined in the introduction, to recover the correct Ewald method for coulomb interaction in two dimensions, one may not take the limit $\eta\rightarrow 0$ in Eqs.(\ref{E_eta_RPM_2D},\ref{E_eta_OCP_2D}), but one must use Eq.(\ref{PowL_log}). For coulomb interactions ($\eta=1$), monolayers of the restricted primitive model  of electrolytes \cite{Weis:98} and of the one component plasma model \cite{Totsuji:78} have been studied previously.\\
In the following, we report some preliminaries numerical results obtained by Monte-Carlo simulations in the canonical ensemble (NAT) \cite{Allen:book:87}. For $\eta$-RPM, the computations are done in a square box with a fixed shape, while for $\eta$-OCP monolayer models the surface $A$ of the simulation box is fixed, but the shape of the basic cell is allowed to fluctuate \cite{Weis:01}. Periodic boundary conditions and the minimum image convention are applied. For both models, average energies are computed from Eqs.(\ref{E_eta_RPM_2D},\ref{E_eta_OCP_2D}), pair correlations functions are also exactly computed with the same definitions and methods as in refs.\cite{Totsuji:78,Allen:book:87,Weis:98,Weis:01,Mazars:08}. Voronoi constructions and cells for the $\eta$-OCP monolayers are exactly computed as in ref.\cite{Mazars:08}.\\
In Table \ref{TabMC_Powlaw_4}, we report some average energies computed with Monte-Carlo simulations  for the $\eta$-RPM and $\eta$-OCP models. The values obtained with $\eta=1.0$ do well agree with previous results obtained in refs.\cite{Weis:98,Totsuji:78}.\\
On Figure \ref{Fig_PowLaw_1}, we show some snapshots for $\eta$-RPM monolayers (a-b) and $\eta$-OCP models (c-d). Snapshots with $\eta=1.0$ agree with the structure found in systems with coulomb interactions \cite{Totsuji:78,Weis:98}, while for $\eta> 1.0$ short ranged order is more marked. This finding is enforced with the shape of pair correlation functions given in Figure \ref{Fig_PowLaw_2}.\\
On Figure \ref{Fig_PowLaw_2} (a), pair correlation functions in the $\eta$-RPM monolayers between hard spheres that carry positive charge are represented. For $\eta=1.0$, $g_{++}(r)$ is exactly the same as the one obtained for the standard RPM model reported on FIG.10(a) of ref.\cite{Weis:98}. As already shown on the snapshot of Figure \ref{Fig_PowLaw_1}, when $\eta>1$ the short ranged order and correlation between particles are more pronounced than with coulomb interaction ; they are less if $\eta < 1$. A similar behavior is found in $\eta$-OCP monolayers ; on Figure \ref{Fig_PowLaw_2}(b), we report pair correlation functions between point particles in $\eta$-OCP monolayers. For $\eta=1$, $g(r)$ is exactly the same as in FIG.2 of ref.\cite{Totsuji:78}.

\section{Discussion.}

From the potential $\phi_\eta(\bm{r})$ computed with the Ewald method for tridimensional, quasi-two dimensional or two dimensional systems derived in section 2 and 3, one may easily obtain any physical quantity related to the potential in a similar way as it is done for the energy.\\
In ref.\cite{Johnson:07}, Johnson and Ranganathan have proposed a generalized approach to Ewald sums for diverse long ranged central potentials, including inverse power law potential as $\phi(r)=r^{-(1+\delta)}$ with $0<\delta<1$. Their derivation is based on the split of the interaction potential as (cf.Eq.(6) of ref.\cite{Johnson:07})
\begin{equation}
\label{deco_Johnson}
\hspace{-0.6in}\displaystyle \phi(r)=(1-f(r))\phi(r)+f(r)\phi(r)=\left(1-\mbox{erf}(\alpha r^\mu)\right)\phi(r)+\mbox{erf}(\alpha r^\mu)\phi(r)
\end{equation}
where $\alpha$ and $\mu$ are chosen conveniently. In lattice sums, the first contribution in the right hand side of Eq.(\ref{deco_Johnson}) is evaluated as the real space contribution and the second as the reciprocal space contribution by using the Fourier transform of $f(r)$.\\ 
In the present paper, the derivation of Ewald methods for inverse power law interactions is done by using the Poisson-Jacobi identities ; according to the results for tridimensional systems in Eqs.(\ref{PowL_4},\ref{Cont_eta}), it corresponds to a choice of the screening function as $f(r)=\gamma(\eta/2,\alpha^2 r^2)/\Gamma(\eta/2)$. For coulomb potential ($\eta=1$, or $\delta=0$ in notations of ref.\cite{Johnson:07}) both choices are strictly equivalent, for other values of $\eta$ formulas differ since the choice of the screening of pseudo-charges differ ; however, both derivations should lead to the same numerical results \cite{Fortuin:77,Rhee:89,Lee:97}.\\
The preliminary numerical results for $\eta$-RPM and $\eta$-OCP monolayers, given in section 4., show that the Coulomb potential case is correctly reproduced by inverse power law interactions with $\eta=1$ when compared to previous results \cite{Totsuji:78,Weis:98}. A longer numerical study of these systems is ongoing.

\section*{Acknowledgments}
It is a pleasure for me to thank Prof. Sabine Klapp and Jean-Jacques Weis for interesting discussions on Inverse Power Law potentials. This work was granted access to the HPC resources of IDRIS under the allocation 2010092104 made by GENCI (Grand Equipement National de Calcul Intensif).

\end{document}